\documentclass[pre,reprint,superscriptaddress]{revtex4-1}
\usepackage{graphicx}
\usepackage{xcolor}
\usepackage{amsmath}
\usepackage{amssymb}
\usepackage[caption=false]{subfig}
\usepackage{gensymb}
\usepackage{appendix}
\usepackage{hyperref}

\newcommand{\red}{\textcolor{red}}

\begin{document}

\title{Non-equilibrium dynamics of disordered fractal spring network with active forces}

\author{Debjyoti Majumdar}
\email{debjyoti@post.bgu.ac.il}
\affiliation{Alexandre Yersin Department of Solar Energy and Environmental Physics, Jacob Blaustein Institutes for Desert Research,\\
Ben-Gurion University of the Negev, Sede Boqer Campus 84990, Israel}

\author{Sadhana Singh}
\email{sadhana@post.bgu.ac.il}

\author{Rony Granek}
\email{rgranek@bgu.ac.il}
\affiliation{Avram and Stella Goldstein-Goren Department of Biotechnology Engineering,\\
and Ilse Katz Institute for Nanoscale Science and Technology,\\
Ben-Gurion University of The Negev, Beer Sheva 84105, Israel}

\date{\today}

\begin{abstract}
We investigate the non-equilibrium dynamics of active bead-spring critical percolation clusters under the action of monopolar and dipolar forces. Previously, Langevin dynamics simulations of Rouse-type dynamics were performed on a deterministic fractal -- the Sierpinski gasket -- and combined with analytical theory [Chaos {\bf 34}, 113107 (2024)]. To study disordered fractals, we use here the critical (bond) percolation infinite cluster of square and triangular lattices, where beads (occupying nodes) are connected by harmonic springs. Two types of active stochastic forces, modeled as random telegraph processes, are considered: force monopoles, acting on individual nodes in random directions, and force dipoles, where extensile or contractile forces act between pairs of nodes, forming dipole links. A dynamical steady state is reached where the network is dynamically swelled for force monopoles. The time-averaged mean square displacement (MSD) shows sub-diffusive behavior at intermediate times longer than the force correlation time, whose anomalous exponent is solely controlled by the spectral dimension $(d_s)$ of the fractal network yielding MSD $\sim t^{\nu}$, with $\nu=1-\frac{d_s}{2}$, similar to the thermal system and in accord with the general analytic theory. In contrast, dipolar forces require a diverging time to reach a steady state, depending on the fraction of dipoles, and lead to network shrinkage. Within a quasi-steady-state assumption, we find a saturation behavior at the same temporal regime. Thereafter, a second ballistic-like rise is observed for networks with a low fraction of dipole forces, followed by a linear, diffusive increase. The second ballistic rise is, however, absent in networks fully occupied with force dipoles. These two behaviors are argued to result from local rotations of nodes, which are either persistent or fluctuating. We further extend our study of dipolar forces to dilution regimes above the isostatic, rigidity percolation threshold, where weak forces effectively do not shrink the network in a steady state. Instead, for the triangular lattice, an incipient discontinuous collapse transition occurs above a critical force amplitude value. We find a continuous crossover to a collapsed state for the non-diluted square lattice, i.e., marginally stable. We suggest that active disordered solids be poised above the isostatic point to be stable against active dipolar forces, even if they are very weak and rare. 
\end{abstract}

\maketitle

\section{Introduction \label{SC1}}

Networks provide a generalized framework to understand natural phenomena across various length scales. Recently, the behavior of networks modulated by active noise has gained much interest, mainly due to its relevance in biological and soft-matter systems. Popular examples include bio-polymer networks with molecular motors such as the networks of filamentous proteins forming {\it cytoskeleton}, in particular the `actomyosin' system that is composed of the actin network with embedded myosin motor proteins that generate internal stresses. More generally, active gels have become a focal point of interest as a subfield of active matter \cite{mac2008, alvarado2013, prost2015, alvarado2017}. Other examples from the field of soft matter where the network picture is prevalent include granular packings, foams near the jamming transition, various fibrous networks \cite{broedersz2011, dasbiswas2023}, and colloidal suspensions. What is common in most of these examples is that the systems are inherently out of equilibrium with active non-equilibrium fluctuations, capable of modulating the network properties, such as its stiffness.

A network's properties generally depend on the degree of connectivity among the spanning nodes, resulting in different regimes of functional behavior. These regimes are controlled by fixed points around which the system shows different characteristics depending upon which side the system is poised at. Fixed points concerning connectivity include the ``percolation" point, where the system is critical with a diverging correlation length and a single large cluster spanning in every dimension, showing fractality, emerges. Above the percolation point, for nodes connected with central-force spring-like bonds, lies the ``isostatic" point, where rigidity percolates throughout the system, making it ``just" (marginally) stable. Each of these regimes holds special relevance in biology, e.g., the fractal regime could be relevant to understanding the compact chromatin, whose fractal dimension was found to be $d_f=2.4 (<3)$ through small angle neutron scattering \cite{metze2013}. At the same time, the mechanically rigid region could be relevant for studying the actin-myosin network, which at short time scales behaves as an elastic material capable of sustaining and transmitting mechanical stresses \cite{gnesotto2019,dasbiswas2023}. Although several studies exist on the mechanical behavior of marginally stable network systems for both passive and active systems \cite{gnesotto2019} and references therein, the fractal regime remains less explored, with only a few works focused on the equilibrium aspects of the network \cite{jurjiu2003,jurjiu2018, grosberg1993, tamm2015}, however, the effect of active forces was not addressed in those studies. Only recently, Singh and Granek \cite{singh2024} considered the non-equilibrium effects due to active forces on the Sierpinski gasket.

In the actomyosin network, one of the important phenomenon observed {\it in-vivo} is the cytoplasmic rotational flow in mammalian cells \cite{woodhouse2012, suzuki2017}. It has been proposed that the myosin activity in the actomyosin system is responsible for this rotation \cite{kumar2014}. The myosin motors are known to produce mainly dipolar forces on the actin filaments. However, these internal forces do not exert a net torque on the system and, therefore, cannot create simple rigid body rotations. Theoretical continuous active gel theory was put forward to explain these rotations \cite{kumar2014}. Persistent ``crawling" rotations were also found from the theoretical treatment of an active fractal network with stochastic activity of force dipoles for random spatial distribution of dipolar forces and explained as deformable body rotations due to spontaneous symmetry breaking of the force dipole spatial distribution \cite{singh2024}. In order to check the hypothesis that myosin dipolar forces are, in fact, inducing the cytoskeleton rotations, it is required to check the conditions under which rotations exist in non-fractal networks as well, that resemble the actomyosin system. In the latter, the long actin filaments are entangled and partially crosslinked, e.g., by passive $\alpha$-actinin, and the myosin motor proteins act as sources of force dipoles.

In this paper, therefore, we will investigate the activity-induced effects on the mechanical and dynamical properties of minimal network models, such as the square lattice, leveraging the simplicity of these models for theoretical and computational treatment. Two kinds of active forces -- monopoles and dipoles -- will be considered, along with the thermal system, for comparison. Later, we extend our study to fully connected and sparsely diluted mechanically rigid triangular networks as well, to understand the dynamical collapse and rotation under the action of dipole forces. Thereby, a quantitative understanding of how the interplay of active forces and network connectivity facilitates a collapse transition and bulk rotation, is developed.

The rest of the paper is organized in the following manner: in Sec. \ref{mini_review}, we briefly review the concepts of fractality. In Sec. \ref{model}, we discuss the disordered fractal model, the distribution of active forces, and the parameter values connected to real physical systems. In Sec \ref{simulation}, we describe the numerical simulation details, such as the integrator scheme used, the observables of interest, and time averaging. In Sec. \ref{analytical}, we briefly review the exact analytical results obtained for this model in Ref. \cite{singh2024}. In Sec. \ref{thermal}, we describe the results for a passive thermal network, and in Sec. \ref{monopole}, we discuss monopolar active forces and active dipolar forces in Sec. \ref{dipole}. Finally, in Sec. \ref{conclusion}, we conclude the paper with a comparative discussion and possible future outlooks.


\section{Fractals - mini review \label{mini_review}}

Before proceeding, let us briefly recall the properties of a fractal. A system exhibits fractality when its structural connectivity manifests scale-invariance and self-similarity. Fractals, in general, are characterized by a few broken dimensions \cite{stauffer1992}: (i) the mass fractal dimension $d_f$ following the scaling $M(r)\sim r^{d_f}$ of the mass $M(r)$ enclosed in concentric circles of radius $r$, (ii) the spectral dimension $d_s$ governing the scaling $g(\omega)\sim \omega^{d_s-1}$ of the vibrational density of states $g(\omega)$ with frequency $\omega$ \cite{alex1982,alex1989}, and the topological dimension $d_l$ that governs the scaling $M(l)\sim l^{d_l}$ of the mass enclosed in concentric ``spheres" of radius $l$ in the topological space. The three broken dimensions obey the inequalities: $1\leq d_s \leq d_l \leq d_f \leq d$, where $d$ is the dimension of the embedding Euclidean space. The fractal dimension $(d_f)$ provides information regarding the arrangements of the fractal nodes in space, and the spectral dimension $(d_s)$ relates to the density of the vibrational states.

Again, fractals can be broadly classified into two different classes: deterministic and disordered fractals. For example, the Sierpinski gasket is a deterministic fractal created by successively joining three motifs of $n-1$th generation to obtain the $n$th generation fractal. The advantage is that it guarantees the existence of a scale-free structure and exact analytical solution of different quantities. On the other hand, an example of a disordered fractal would be the percolating infinite cluster backbone at the bond percolation threshold on a two-dimensional square lattice. 

\section{The model  \label{model}}

\begin{figure}[t]
\centering
\includegraphics[width=\linewidth]{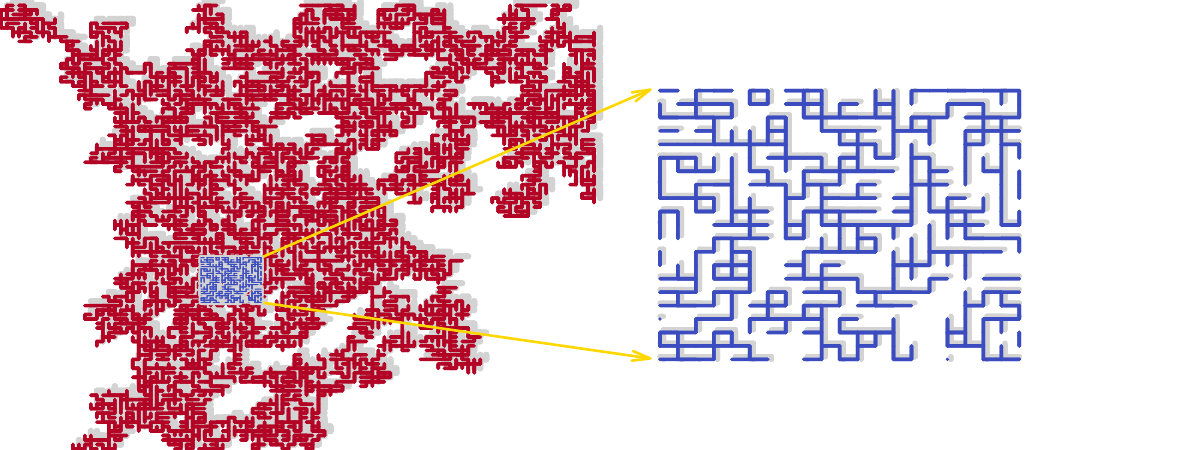}
\caption{\textbf{Fractal lattice.} A disordered fractal constructed from the infinite cluster close to the bond percolation critical point $(p_c)$ of the square lattice at $p=p_c+10^{-3}$.}
\label{fig_fractallattice}
\end{figure}

To model the disordered fractal network, we use the critical infinite cluster at the two-dimensional bond-percolation threshold on a square lattice [Fig. \ref{fig_fractallattice}], specifically at $p=p_c + 10^{-3}$,  where $p$ is the probability with which a bond exists and $p_c=1/2$ is the bond-percolation critical point. Also considered in the latter part are diluted triangular lattices for which $p_c=0.347$. The lattice sites represent beads which are connected via harmonic springs of rest length $b$ and spring constant $m\omega_0^2$. The system evolves following the Hamiltonian 
\begin{equation}
\label{hamiltonian}
H=\frac{1}{2}m\omega_0^2\sum_{\langle ij\rangle}(\vec{r}_i - \vec{r}_j - b \hat{r}_{ij})^2
\end{equation}
which $\langle ij \rangle$ denotes pair of neighboring sites connected by springs of self-frequency $\omega_0$, $m$ is the bead mass. In the $\lim b\rightarrow 0$, Eq. \ref{hamiltonian} reduces to the scalar elasticity Hamiltonian \cite{singh2024}. For simulations we have used $b=1$ unless specified otherwise. Also, we have not considered the effects of excluded volume or hydrodynamic interactions between the nodes in our model.


We consider two classes of stochastic active forces -- force monopoles and force dipoles -- acting on the nodes, distributed randomly and uniformly over the network. Force monopoles are given a random orientation $\theta_i\in [0,2\pi]$, which is fixed at the beginning of the simulation, and the force field is given by 
\begin{equation}
\vec F(\vec r,t) = \sum_j \vec f_j(t) \delta(\vec r - \vec r_j),
\end{equation}
where $\vec f_j(t)$ is the time-dependent active force acting on $j$th bead at time $t$ and the $\delta-$function ensures that the force originates right at the node. Unlike monopoles, the force field for dipoles is given by 
\begin{equation}
\vec F(\vec r,t) = \sum_j \vec f_j(t) [ \delta(\vec r - \vec r_j) - \delta(\vec r - \vec r_j-\vec \epsilon_j)],
\end{equation}
where $\vec \epsilon_j~(\mid \vec \epsilon_j \mid = b)$ is a randomly chosen vector directed along one of its nearest neighbors. A positive (or negative) $\vec f_j$ implies an inward contractile (or extensile) force. 

The evolution of these active forces follow the random telegraphic process \cite{gardiner1991}, where the random variable $\vec f_j(t)$ stochastically fluctuates between two values: $0$ for OFF-state, and $f$ for ON-state \cite{gardiner1991}. At $t=0$ the probability for the active sources to be in an ON (or OFF) state is given by $\mathcal{P}$ (or $1-\mathcal{P}$). Each individual sources of active force undergo a correlated dynamics, of which the transition probabilities are  discussed in Appendix \ref{tlnoise}. Note that the active forces are only temporally correlated without any spatial correlation. The autocorrelation function of the active forces can be written as \cite{singh2024}
\begin{equation}
\langle \vec f_i(t) \cdot \vec f_j(0) \rangle = f^2 \mathcal{P}^2 \hat f_i \cdot \hat f_j + f^2\delta_{ij} \mathcal{P}(1-\mathcal{P})e^{-t/\tau},
\end{equation}
where $t$ is the lag time and $\tau$ is the force decorrelation time. The ON and OFF probabilities are to be identified, e.g., with the molecular motors' ``attachment" and ``detachment" rates within a biopolymer network. For simplicity, we have chosen $\mathcal{P}=1/2$ throughout the paper.


\section{Numerical simulation and Observables \label{simulation}}

\subsection{Integration scheme}

We perform Langevin dynamics (LD) simulations of the bead-spring disordered fractal network in the over-damped limit, i.e., $\frac{d \vec{v}}{dt}\rightarrow 0$, thereby neglecting any inertial effects. For simplicity, we have used friction coefficient $\gamma=1$. To numerically integrate the equations of motion, we use the two-point finite difference method also known as the Euler-Maruyama method. See discussion in Appendix \ref{simulation_appendix}. Our simulations use the following unit system; position $\vec r$ is mentioned in units of bond length $b$, active force $\vec f$ in units of $m\omega_0^2b$, time $t$ and $\tau$ are in units of $\tau_0=\gamma/(m\omega_0^2)$ and an integration time step of $\delta t=10^{-2}$. For choice of parameter values refer to Appendix \ref{parameter_choice}.

\begin{figure}[b]
\centering
\includegraphics[width=.47\linewidth]{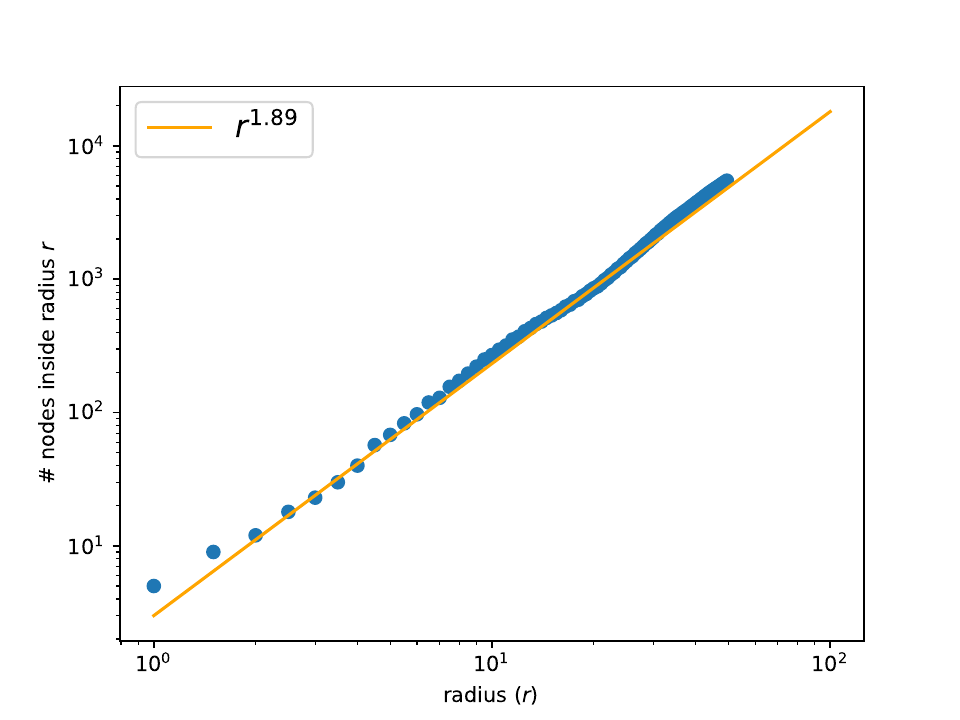}
\includegraphics[width=.47\linewidth]{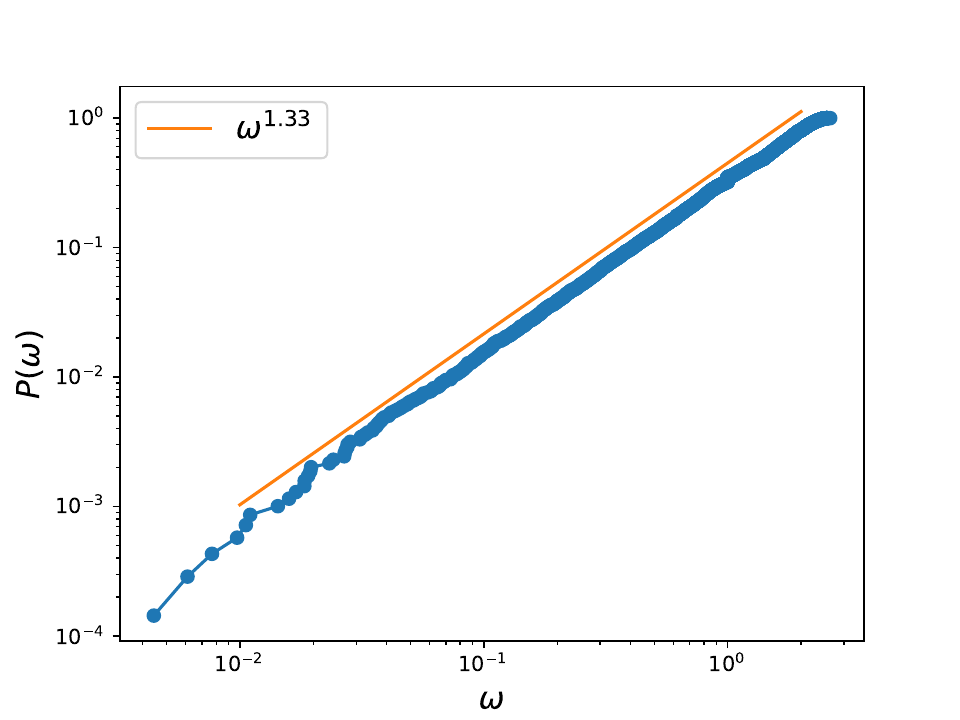}
\caption{\textbf{Fractal dimensions.} (a) Mass fractal dimension $d_f=1.89$, and (b) the cumulative density of vibrational states yielding spectral dimension $d_s=1.33$.}
\label{fig_fractallattice_prop}
\end{figure}
\subsection{Fractal lattice properties}
The infinite clusters of the disordered fractal are indexed, and the infinite cluster is isolated using the Hoshen-Kopelman algorithm \cite{hoshen1976} for square lattice and a depth-first search in the case of triangular lattice. The correlation length around the percolation transition point diverges as $\zeta\sim\mid p-p_c \mid^{-\nu}$ with $\nu=1.33$, which is roughly around $\zeta \sim 10^4$ for our choice of  $p=0.501$, much larger than our lattice dimensions ensuring that we are well inside the fractal regime. To evaluate the spectral dimension $d_s$, we find the eigenvalues ($\lambda$) of the $N\times N$ connectivity matrix $\textbf{C}$ (also called ``Kirchhoff matrix"), where $N$ is the number of nodes in the infinite cluster. The eigenvalues $\lambda$'s are related to the eigen frequencies (or normal modes )  $\omega_\alpha$ as, $\omega_\alpha=\sqrt\lambda_\alpha$. For fractals, the density of states scales as $g(\omega)\sim\omega^{d_s-1}$, therefore, the cumulative density of states $P(\omega) \sim \int_0^{\omega} g(\omega) d\omega \sim \omega^{d_s}$. For our infinite cluster, we found $d_f=1.89$ and $d_s=1.33$ [Fig. \ref{fig_fractallattice_prop}] in good agreement with the already established values \cite{stauffer1992}.

\subsection{Observables}
We looked at the time-averaged mean squared displacement (MSD) denoted by $\triangle \vec r^2$ to investigate the dynamical behavior. Since we are working with disordered lattices, the MSD was time averaged over $n$ randomly chosen nodes which can be written as
\begin{equation}
\triangle \vec r^2 = \frac{1}{t_{tot}-t} \sum_{t_i}^{t_{tot}-t} \frac{1}{n} \sum_{j=1}^{n} \left( \vec{r}_j(t_i+t) - \vec{r}_j(t) \right)^2. 
\label{MSD1}
\end{equation}
We found the result to be invariant to the order in which the average is taken over time and number of nodes. To calculate the MSD, we store the positions of $n$ nodes at consecutive integration steps. To correct any drift in the center of mass, we need to subtract the average displacement in the same time duration, and Eq. \ref{MSD1} needs to be modified as  
\begin{equation}
\triangle \vec r^2 = \frac{1}{t_{tot}-t} \sum_{t_i}^{t_{tot}-t} \frac{1}{n} \sum_{j=1}^{n} \left( \vec{r}_j(t_i+t) - \vec{r}_j(t_i) - \langle\vec{r}(t)\rangle \right)^2
\label{MSD2}
\end{equation}
where $\langle\vec{r}(t)\rangle$ is the average displacement in time $t$ and is given by 
\begin{equation}
\langle\vec{r}(t)\rangle = \frac{1}{t_{tot}-t} \sum_{t_i}^{t_{tot}-t} \frac{1}{n} \sum_{j=1}^{n} \left( \vec{r}_j(t_i+t) - \vec{r}_j(t_i) \right).
\label{MSD3}
\end{equation}
Besides MSD, to study the collapse transition, we also looked at the network radius of gyration $R_g$ given by,
\begin{equation}
R_g = \sqrt{\frac{1}{N} \sum_{i=1}^{N} \mid \vec r_i - \vec r_{CM} \mid^2},
\end{equation}
where $\vec r_{CM}=\frac{1}{N}\sum_{i=1}^N \vec{r}_i$ is the fractal center of mass and $N$ is the total number of nodes.

\section{ Analytical results \label{analytical}}

The analytical theory for this model has been developed in Ref. \cite{singh2024}. The thermal and active mean squared displacement ($\triangle \vec{r}^2$) can be expressed in terms of the normal modes and upon time averaging one can write 
\begin{equation}
\langle (\vec r(t) - \vec r(0))^2 \rangle = \langle \triangle \vec r(t)^2 \rangle_{th} + \langle \triangle \vec r(t)^2 \rangle_{ac}    
\end{equation}
as the sum of the individual thermal and active contribution where,
\begin{eqnarray}
\langle  \triangle \vec r(t)^2 \rangle_{th} = \frac{1}{N} \sum_{\alpha} \frac{2dk_BT}{m \omega_{\alpha}^2} \left( 1-e^{-\Gamma_{\alpha}t} \right) ,\label{eqthex} \\
\langle  \triangle \vec r(t)^2 \rangle_{ac} = \frac{1}{N} \sum_{\alpha} \frac{2\phi f^2 p(1-p)W_{\alpha}\Lambda_{\alpha}^2}{\Gamma_{\alpha}(\Gamma_{\alpha} + \tau^{-1})} \label{eqacex} \\
 \times \left( 1+ \frac{\tau^{-1}}{\Gamma_{\alpha}-\tau^{-1}}e^{-\Gamma_{\alpha}t} - \frac{\Gamma_{\alpha}}{\Gamma_{\alpha}-\tau^{-1}}e^{-t/\tau}  \right)   \nonumber
\end{eqnarray}

obtained by summing over all vibrational states $(\omega_{\alpha})$. The combined active system has three timescales: first, the force decorrelation time $(\tau)$, and the other two inherent to the fractal network $(\tau_0,\tau_N)$ giving the shortest and the longest network relaxation times, given by $\tau_0=\gamma/(m\omega_0^2)$ and $\tau_N\sim \Gamma(\omega_{min})^{-1}\sim\tau_0 N^{2/d_s}$.

For the thermal system it has already been shown in Refs. \cite{granek2005, granek2011, reuveni2012, reuveni2012p2} 

\begin{equation}
    \langle  \triangle \vec r(t)^2 \rangle_{th} = B_{th} t^{\nu_{th}},
\end{equation}
where $\nu_{th} = 1-\frac{d_s}{2}$, provided $d_s<2$, and the prefactor $B_{th} \sim dk_BT\gamma^{d_s/2-1}$.

In active systems, for short times $t<<\tau<<\tau_N$, the theory finds super-diffusive ballistic scaling behavior
\begin{equation}
\langle \triangle \vec{r}(t)^2  \rangle_{ac} = B_{acs} t^2, ~~~~~~ t<<\tau,
\end{equation}
where the amplitude $B_{acs}$ follows different scaling for force monopoles and dipoles. For times much longer than the force correlation but shorter than the network relaxation time such that $\tau<<t<<\tau_N$, we have for force monopoles 
\begin{equation}
\langle \triangle \vec{r}(t)^2  \rangle_{ac} = B_{ac} t^{\nu_{ac}}, ~~~~~~ t<<\tau,
\end{equation}
where $\nu_{ac}=1-d_s/2$ that is identical to the thermal subdiffusion exponent and $B_{ac}=\phi f_0^2 \tau$. Further, since $1-d_s/2-d_s/d_l<0$ for force dipoles, the theory predicts a saturated MSD, with the absence of any anomalous subdiffusion regime.

\section{Results}

\subsection{Thermal network \label{thermal}}
\begin{figure}[t]
\centering
\includegraphics[width=\linewidth]{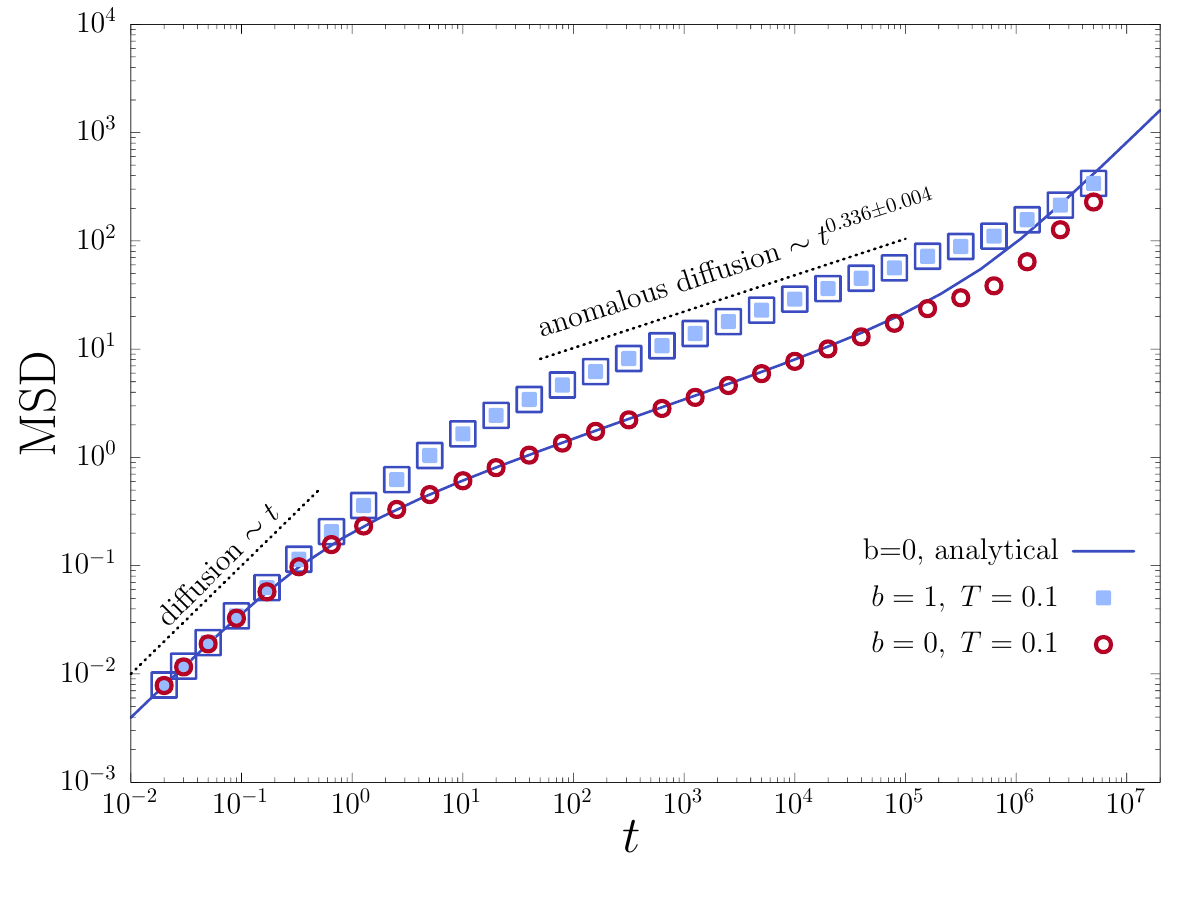}
\caption{\textbf{Passive-thermal network.} Log-log plot of MSD of a passive-thermal network at $T=0.1$ and averaged over $50$ nodes. The data points correspond to numerical simulations of networks with spring rest length $b=0$ and $b=1$, while the continuous solid line is the exact MSD for $b=0$ case, calculated using Eq. \ref{eqthex}. The three scaling regimes are: initially $\triangle r^2 \sim t$, in the intermediate regime it scales scale as $\triangle r^2 \sim t^{0.336}$, and again scales linearly in the long time scale $t >\tau_N$.}
\label{fig_msd_thermal}
\end{figure}
First, we investigate the effect of fractality on the dynamics of a passive network at temperature $T=0.1$, driven only by thermal fluctuations. Two cases are considered, for spring rest length $b=0$ and $b=1$. The system size decreases while reaching the dynamical steady state as found from the $R_g$. Once the steady state is reached, we start gathering data at every step to calculate the time-averaged MSD as shown in Fig. \ref{fig_msd_thermal}. We averaged MSD over 50 randomly chosen nodes across the network. The MSD follows a linear scaling $\triangle r^2\sim t$ in the short $(t<<\tau_0)$ and long $(t>>\tau_N)$ timescales. In the intermediate regime, $\tau_0<t<\tau_N$ MSD follows a sub-diffusive scaling whose exponent is governed by the  spectral dimension $d_s$ of the fractal network given by $\nu_{th} = 1-\frac{d_s}{2}=0.33(5)$ for $d_s=1.33$ (from Fig.~\ref{fig_fractallattice_prop}). Fitting over the range $t\in[10^2,10^5]$ we obtain $\nu=0.336\pm0.004$, in close agreement with the exact value $\nu_{th}$. The short time represents independent nodes diffusion while the long time represents center of mass (CM) diffusion without any drift. The analytically exact MSD for spring rest length $b=0$ is found from the normal modes of the fractal network using Eq. \ref{eqthex}. To study the effect of number of averaging nodes, we additionally calculate MSD only from a single node. In that case, while the initial and long-time MSD shows linear scaling, the intermediate regime shows fluctuations, suggesting that the scaling in the intermediate regime is the most sensitive to the underlying disordered fractal structure and, thereby upon the number of averaging nodes. 

\subsection{Force monopoles \label{monopole}}

For force monopoles, the directions $(\{\theta_i\})$ of the active forces are assigned randomly from a uniform distribution of $\theta_i \in \mathcal{U}(0, 2\pi)$ ``for once" at the start of the simulations. We consider all beads $(\phi=1)$ to be source of monopole. Both athermal and thermal cases are considered; for the athermal $(T=0)$ case we consider $\tau=1$ and $100$. To see the effect of thermal noise we simulate at an elevated temperature of $T=0.1$. Initially, at $t=0$, the individual monopole sources are in an ON (or OFF) state with a probability $\mathcal{P}$ (or $1-\mathcal{P}$), which we have chosen to be $\mathcal{P}=1/2$ for all cases.

\begin{figure}[t]
\centering
\hspace{-7cm}(a)\\
\includegraphics[width=\linewidth]{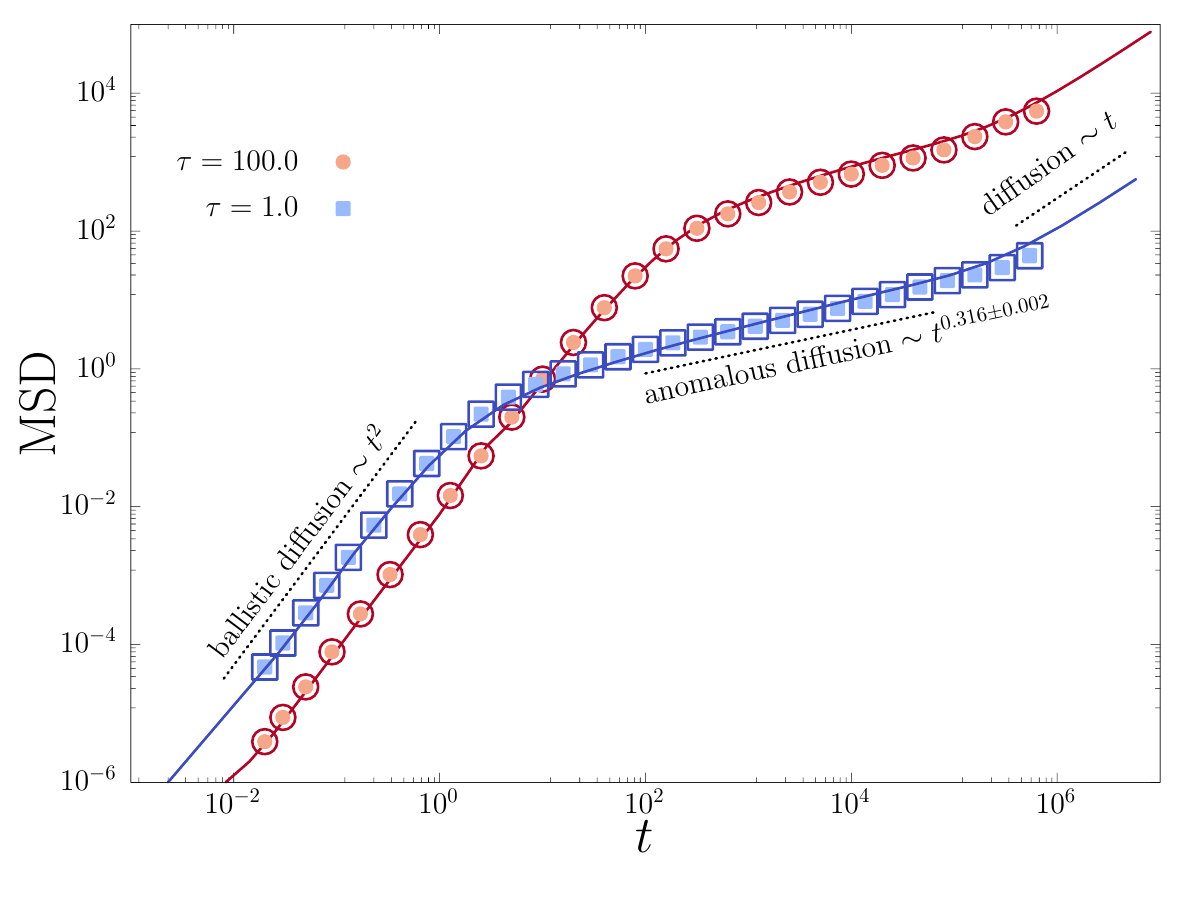}\\
\hspace{-3cm}(b)~~~~~~~~~~~~~~~~~~~~~~~~~~~~~~~~~(c)\\
\includegraphics[width=.49\linewidth]{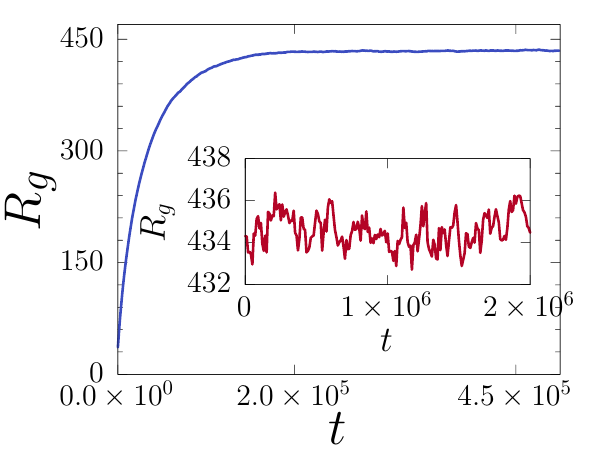}
\includegraphics[width=.49\linewidth]{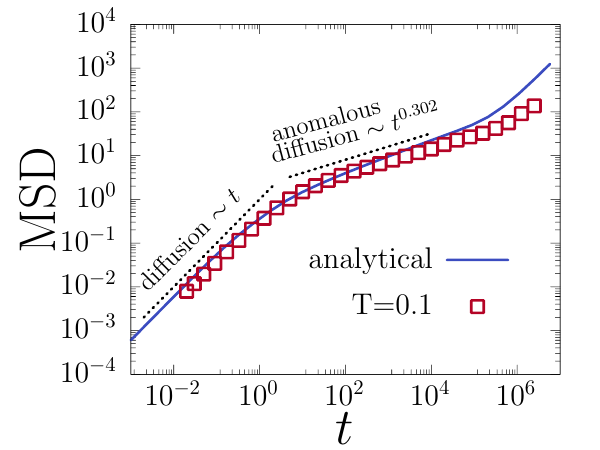}
\caption{\textbf{Active network with force monopoles:} (a) Log-log plot of MSD averaged over 50 nodes of a network with a fraction $\phi=1$ of nodes as active monopoles. Two different decorrelation times are considered: $\tau=1$ and $\tau=100$, with force amplitude $f=1$ for both cases. The different scaling regimes are shown using the dotted straight lines. (b) $R_g$ during the initial run to take the system to steady state. Inset showing the fluctuation in $R_g$ during data accumulation for calculating the MSD. (c) MSD with active monopole forces $(f=1)$ at temperature $T=0.1$. The solid lines show the analytically calculated MSD for $b=0$ network using Eqs. \ref{eqthex} and \ref{eqacex}.}
\label{fig_monopole}
\end{figure}

Before we start gathering data to calculate MSD [Fig. \ref{fig_monopole}(a)], we equilibrate the system for $5\times10^5$ time steps while monitoring the system's configurational energy $(U)$ and the radius of gyration $(R_g)$, ensuring that these quantities reach a steady value [see Fig. \ref{fig_monopole}(b)]. Notice that the dynamical steady state is swelled according to the proposed theory \cite{singh2024}. Additionally, we also keep track of the $R_g$ during the MSD calculation and make sure it does not change substantially other than the usual fluctuation about a mean value [see Fig. \ref{fig_monopole}(b) inset].

\textit{Dynamical behavior:} In Fig. \ref{fig_monopole}(a), we show the MSD of the fractal network at $T=0$ under the action of force monopoles, averaged over 50 randomly chosen nodes. Two activity timescales are considered: $\tau=1$ and $\tau=100$, with the same force amplitude $f=1$ in both cases. The short-time scaling is ballistic $\triangle \vec r^2 \sim t^2$, followed by a crossover at $t=\tau$ to a sub-diffusive regime $\triangle \vec r^2 \sim t^{\nu}$ with an exponent $\nu=0.31633\pm 0.002541$ at intermediate times, and later succeeded by a second ballistic regime (not shown) due to an incomplete cancellation of the forces for finite systems leading to a CM drift. This effect, however, should vanish in the thermodynamic limit $\lim_{N\rightarrow \infty} \sum_i \vec f_i=0$. The late time ballistic scaling is replaced by a linear scaling, $\triangle \vec r^2 \sim t$, only after we remove the effect of CM drift using Eq. \ref{MSD2} and Eq. \ref{MSD3}. For the best possible fit in the intermediate region, data points are sampled at frequent time intervals than shown in Fig. \ref{fig_monopole}. We further check the validity of the numerical results against the analytically calculated exact MSD values using the normal modes of the associated network and Eq. \ref{eqacex}, which are plotted using continuous solid lines in Fig. \ref{fig_monopole}(a).


In Fig. \ref{fig_monopole}(c), we show the combined effect of active and thermal forces, using the same network, at $T=0.1$ and $f=1$. Notice how even at this small value of finite temperature, the initial ballistic scaling of MSD is replaced by a linear scaling  $\triangle \vec r^2 \sim t$. The amplitude increases compared to the athermal case; however, the sub-diffusive intermediate region persists. On the long time scale $t>10^5$, the deviation from the exact data could result from the finite data size.


\begin{figure}[t]
\includegraphics[width=.43\linewidth]{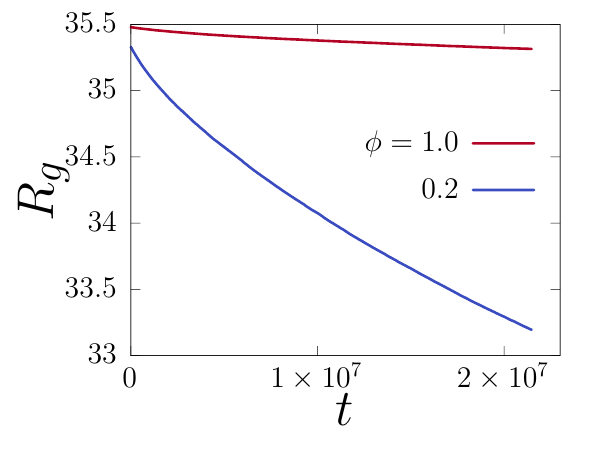}
\includegraphics[width=.55\linewidth]{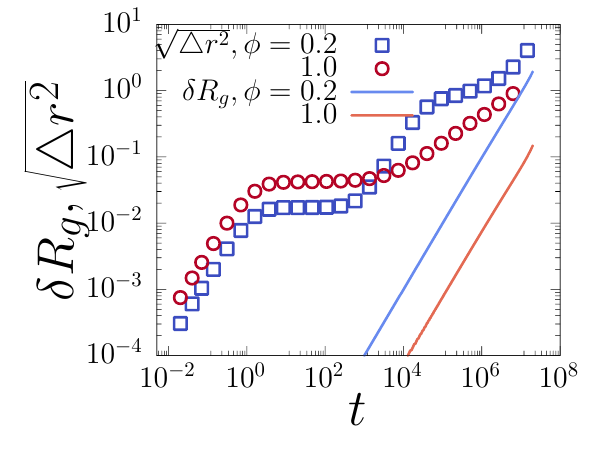}
\caption{\textbf{Pseudo steady state in active force dipole networks.} (a) Variation in $R_g$ during which the data for MSD is gathered. (b) Comparison of change in $R_g$, i.e., $\delta R_g$, during the same time as compared to the $\sqrt{\triangle r^2}$ from plots in (a).} 
\label{fig_rg_dipole}
\end{figure}

\subsection{Force dipoles \label{dipole}}

Next, we come to force dipoles, where force acts between a pair of connected nodes. The forces are either extensile or contractile with no net momentum transfer. To start with, a random fraction $(\phi)$ of links in the infinite cluster are chosen as dipoles. Akin to the monopolar forces, the active force dipoles initially are in an ON or OFF state with a probability $\mathcal{P}$ or $1-\mathcal{P}$, respectively. Here also, we have chosen $\mathcal{P}=1/2$. However, unlike the monopoles, the directions of the dipoles are decided depending on the immediate positions of the nodes while the system evolves. Force dipoles, for example, could arise from molecular motor “pairs” in biological networks. Therefore, it is perhaps appropriate to impose an excluded volume interaction among the motors, by excluding a single node from being part of two dipoles, albeit for low density of dipole links, such as $\phi=0.2$. This condition could introduce a small bias in the distribution, the effect of which should be negligible as long as $\phi$ is low. Such an exclusion, however, is infeasible for higher $\phi$ values. 

\begin{figure}[t]
\centering
\includegraphics[width=\linewidth]{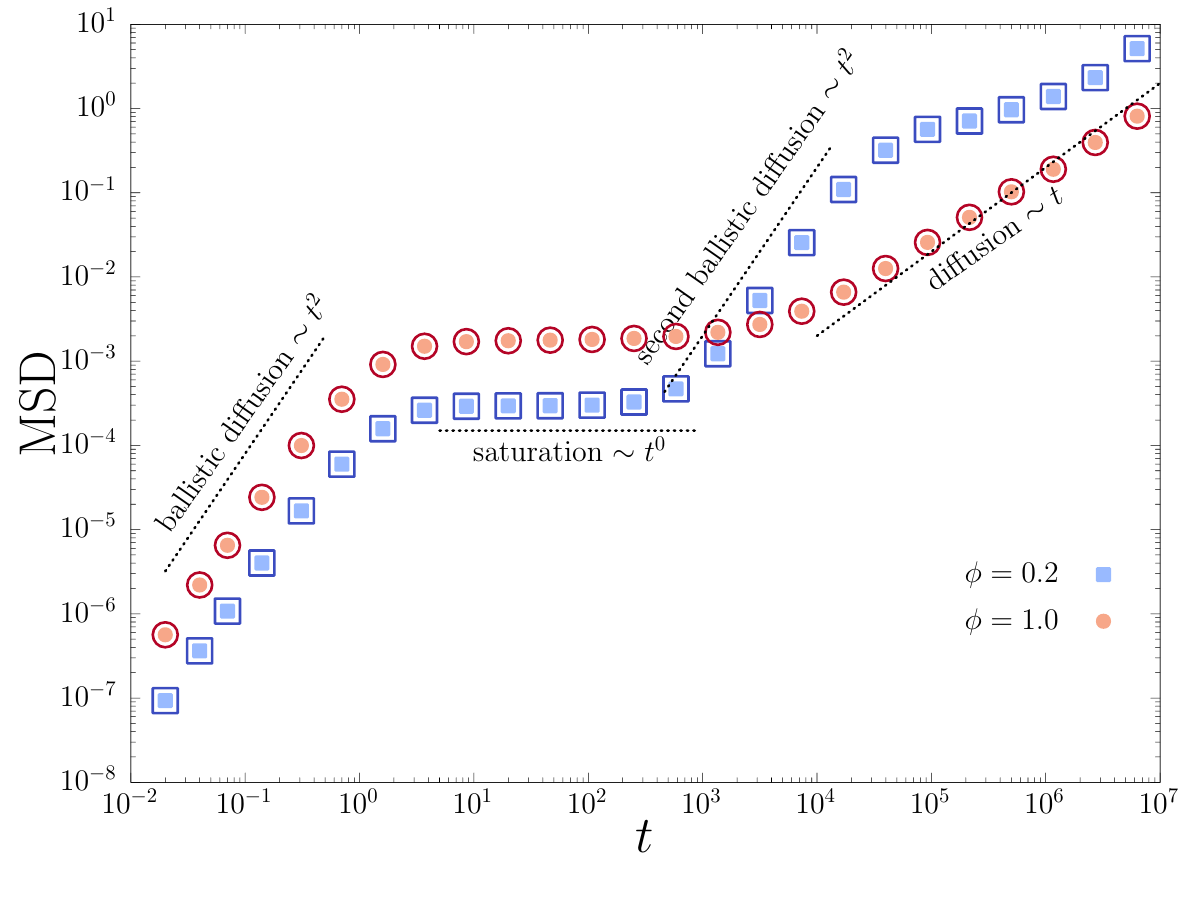}
\caption{\textbf{Active network with force dipoles.} MSD averaged over $10$ random nodes of the fractal network with active force dipoles. Results shown for fraction $\phi=0.2$ and $1$ of dipole links distributed randomly. The active force  amplitude is $f=0.1$ and $\tau=1$ for both cases. The different scaling regimes are shown with dotted straight lines as an guide to the eye.} 
\label{fig_msd_dipole}
\end{figure}

\textit{Dynamical behavior:} 
To start with, we observe that the system did not reach a steady state within our simulation time, but rather displays a slow-continuous decrease ($\sim 2-5\%$) in the $R_g$ value with time, with the $\phi=0.2$ network contracting faster than the $\phi=1$ network [Fig. \ref{fig_rg_dipole}(a)]. This suggests the $t\to\infty$ state of the network is a collapsed structure. However, within a short time interval, the collapse could be small enough to evaluate the time-averaged MSD. The change in $R_g$ $(\delta R_g)$ in Fig. \ref{fig_rg_dipole}(b) at lag times shorter than $~10^6$ are 2 orders of magnitude smaller than the square root time-averaged MSD, which suggests a pseudo-steady state where the MSDs remain meaningful.

In Fig. \ref{fig_msd_dipole}, we show MSD for $\phi=0.2$ and $\phi=1$ fraction of dipole links, with the force amplitude $f=0.1$ and correlation time $\tau=1$ kept same in both cases. For both choices of $\phi$, initially, there is a ballistic-diffusion where MSD scales as $\triangle \vec r^2\sim t^2$, with the $\phi=1$ network having a higher amplitude. This ballistic scaling regime also corresponds to the decorrelation time $\tau$ of the active forces. Following the ballistic regime, a long saturation regime sets in at $t\approx \tau$. At much longer times, a new time-dependent rise appears in the bead MSD, with characteristic dependence on $\phi$, both in terms of scaling exponent and amplitude. For $\phi=1$, we observe a diffusive regime ($\nu=1$), while for $\phi=0.2$ the departure from saturation commences as ballistic-like ($\nu=2$) only later crossing over to a diffusive behavior. Moreover, the MSD amplitude for $\phi=0.2$ is larger than for $\phi=1$. The unexpected ballistic-like behavior for $\phi=0.2$ is associated with effectively free motion of dangling ends (see SM movies), yet the motion of these dangling ends is more restricted when $\phi=1$ thus leading only to a diffusive regime. The early and saturation regimes of the MSDs are similar to those observed in the Sierpinski gasket \cite{singh2024}. The departures from the saturation are, however, different, which are most likely due to the disordered nature of the fractal lattice.



\textit{Collapse:} Next, we examine whether dipole forces lead eventually to the collapse of the critical percolation cluster? We always find, even for extremely low forces, that $R_g$ keeps decreasing for all time, albeit slowly, never reaching a collapsed steady state within our simulation time. This occurs in a force-dependent dynamics, that is, the collapse is faster for higher forces. We infer that for sufficiently long simulations a collapsed steady state would have been reached. In contrast, Singh and Granek \cite{singh2024} found a sharp collapse transition at $f\simeq 1$ for the Sierpinski gasket. To understand this discrepancy, we further analyze the collapse transition in regular undiluted and sparsely diluted lattices. 

Remarkably, even for complete (undiluted) square lattices, the slow network collapse persists (even though the collapse steady state is not observed). This can be rationalized if we recall the Maxwell criterion for stability \cite{comm1}, which, for central pairwise forces, requires $z>2d$ or $z>4$ in 2D, where $z$ is the coordination number, henceforth referred to as the ``isostatic point". Thus, the complete square lattice is only marginally stable to mechanical perturbations.

\begin{figure}[t]
\centering
\hspace{-7cm}(a)\\
\includegraphics[width=\linewidth]{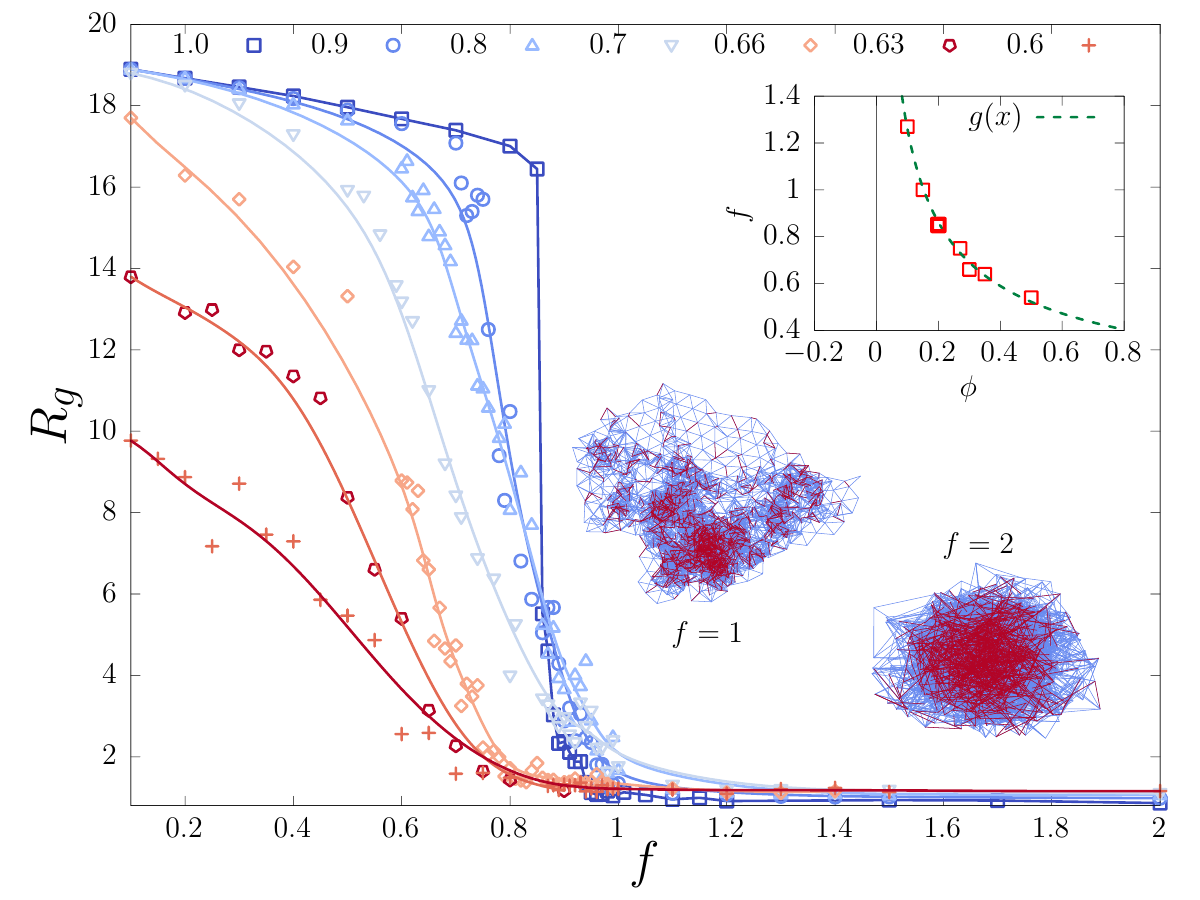}\\
\hspace{-3cm}(b)~~~~~~~~~~~~~~~~~~~~~~~~~~~~~~~~~(c)\\
\includegraphics[width=.49\linewidth]{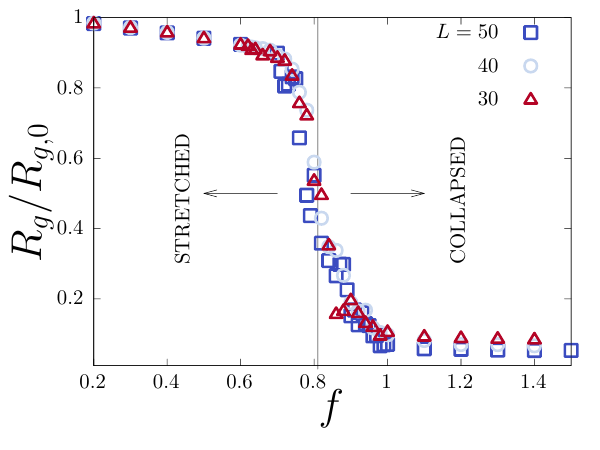}
\includegraphics[width=.49\linewidth]{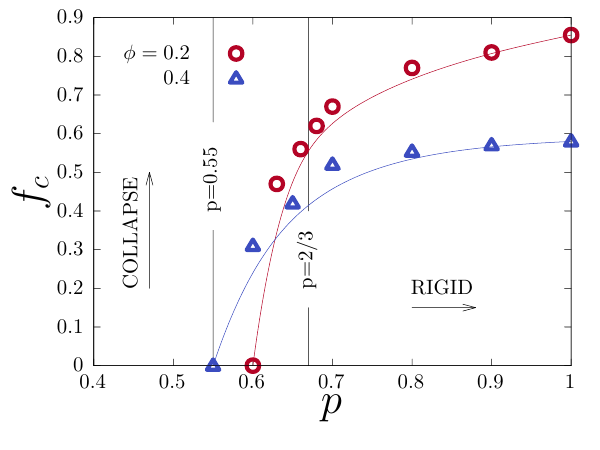}
\caption{\textbf{Triangular lattice collapse under contractile active force dipoles.} (a) Steady state $R_g$ vs. force showing collapse transition of an athermal triangular network made of $50\times50$ nodes, with $\phi=0.2$ fraction of dipole links and different values of dilution $p=1, 0.9, 0.8$ and $0.7$, where $1-p$ is the probability for the dilution of a bond. {Actual mean coordination numbers are $z=5.84, 5.25, 4.66$ and $4.15$, respectively.} The critical force is estimated from the inflection points of the splines approximated from the data. Snapshots of $p=1$ triangular network in the steady state at different force values $f=1$ and $2$ are shown in inset. (Inset) Shows the critical force $f_c$ as a function of $\phi$ for undiluted $(p=1)$ lattice. $g(x)\sim x^{-0.54}$ is a power-law fit to the data points.  (b) $R_g$ in the extended phase scales as the linear dimension ($L$) or  $R_g/R_{g,f=0}$ of the system, shown for $p=0.9$. (c) Critical force $f_c$ as a function of $p$ for $\phi=0.2$ and $0.4$. Here, $f_c=0$ data points are from Eq. \ref{maxwell_modified}. The solid lines are smooth Bezier curve approximation of the data points.}
\label{fig_collapse}
\end{figure}

Hence, for further analysis of the collapse transition, we chose the triangular lattice whose (infinite lattice) mean coordination number is $\langle z\rangle=6$, thereby inherently above the isostatic point $\langle z\rangle_c=4$. In a finite, complete, triangular lattice with non-periodic boundary condition $z=5.84\lesssim 6$. In this case, the system does reach a dynamical steady state where the average $R_g$ is no longer decreasing, yet at small forces the decrease of $R_g$ is minor. A sharp collapse transition is incipient beyond a threshold force $f_c$, e.g., for a $50\times 50$ lattice containing a fraction of $\phi=0.2$ force-dipoles, $f_c\sim 0.9$, as shown in Fig. \ref{fig_collapse}. The critical force depends on $\phi$ and diverges as $\phi\rightarrow 0$, see Fig. \ref{fig_collapse}(a) inset, which implies the absence of a collapse transition as expected. $f_c$ scales with $\phi$ roughly as $f_c\sim \phi^{-0.54}$. The collapse transition remains equally sharp for different $\phi$'s. 


In addition, we considered the effect of dilution for $\phi=0.2$ and $0.4$. Collapse transition for sparsely diluted triangular lattices with $p=0.9$, $0.8$, $0.7$, $0.66$, $0.63$ and $0.6$ are shown in Fig. \ref{fig_collapse}(a) main plot. As we dilute the system, the collapse transition gradually smooths out. In Fig. \ref{fig_collapse}(b), we study the finite size effect and show that $R_g$ scales as $R_{g,0}$, or the system's linear dimension $L$, across the collapse transition, for the $p=0.9$ system. For different system linear sizes $L=30, 40$ and $50$, we do not find any substantial variation in the $f_c$, as we need not require any ad hoc scaling factor along the force axis,  which simply suggests that the estimated $f_c$ should match well for systems in the thermodynamic limit too. Additionally, there are minor fluctuations of $f_c$ associated with different statistical realizations of bond dilution and force-dipole. 

The isostatic point of $\langle z\rangle_c=4$, which is also referred to as ``rigidity percolation" in the literature, corresponds to $p_{rigid}=2/3$ in an infinite triangular lattice. Yet, $f_{c}$ does not vanish at $p=2/3$. In Fig. \ref{fig_collapse}(c), we show how the critical force $f_c$ varies with the dilution parameter $p$. Intriguingly, with the inclusion of dipole forces, a new point for marginal rigidity in the system arises at $p'_{rigid}<2/3$. This change can be understood using the Maxwell's criterion of mechanical stability for systems connected with springs (central force) \cite{comm1}. The inclusion of force dipoles on a fraction $\phi$ of bonds, with a mean fraction $\mathcal{P}$ in their ON-state at any instant, results with increased number of constraints which changes the rigidity percolation point to $z'_c=z_c/(1+\mathcal{P}\phi)$ which means 
\begin{equation}
p'_{rigid}=\frac{p_{rigid}}{(1+\mathcal{P}\phi)}.
\label{maxwell_modified}
\end{equation}
We checked the above formula for two different cases; for $\phi=0.2$ giving $p'=0.606$, and for $\phi=0.4$ giving $p'=0.55$. The extrapolated results from our simulations are in good match with the exact analytical values. Details of the derivation of Eq. \ref{maxwell_modified} are given in Appendix \ref{mmc}. The $f_c=0$ datapoints at $p=0.6$ and $p=0.55$ in Fig. \ref{fig_collapse}(c), corresponding to $\phi=0.2$ and $0.4$ respectively, are obtained using Eq. \ref{maxwell_modified}. 

A similar change in the isostatic point with the inclusion of extra features like bond-bending energy has already been shown in the past \cite{broedersz2011, das2012,dasbiswas2023}. Overall, these observations simply suggest that the collapse as a sharp transition is a cooperative phenomena requiring participation from all parts of the system which can sustain stress, and, therefore, possible only in a mechanically rigid system.  Since the dipolar forces act directly between the nodes, the effect spreads across the network via tension relaxation. Therefore, for systems diluted down to $p<p'$, forces cannot be duly transmitted to the whole network, resulting in a rather gradual reduction in size. Conversely, the extra rigidity in the network itself originates from the same forces. 



\textit{Rotation:} Following the study for the Sierpinski gasket \cite{singh2024}, we also look here for ``crawling" rotations in disordered fractal networks. This may occur, even though force dipoles do not produce net momentum transfer. For a torque-free system, $\sum_i \vec{\Omega}_i \mid\vec{R}_i \mid^2=0$, yet 
 it is possible that $\sum_i \vec{\Omega}_i \neq 0$, where $\vec{\Omega}_i$ is the angular velocity of the $i$th bead. 
 
\begin{figure}[t]
\centering
\includegraphics[width=\linewidth]{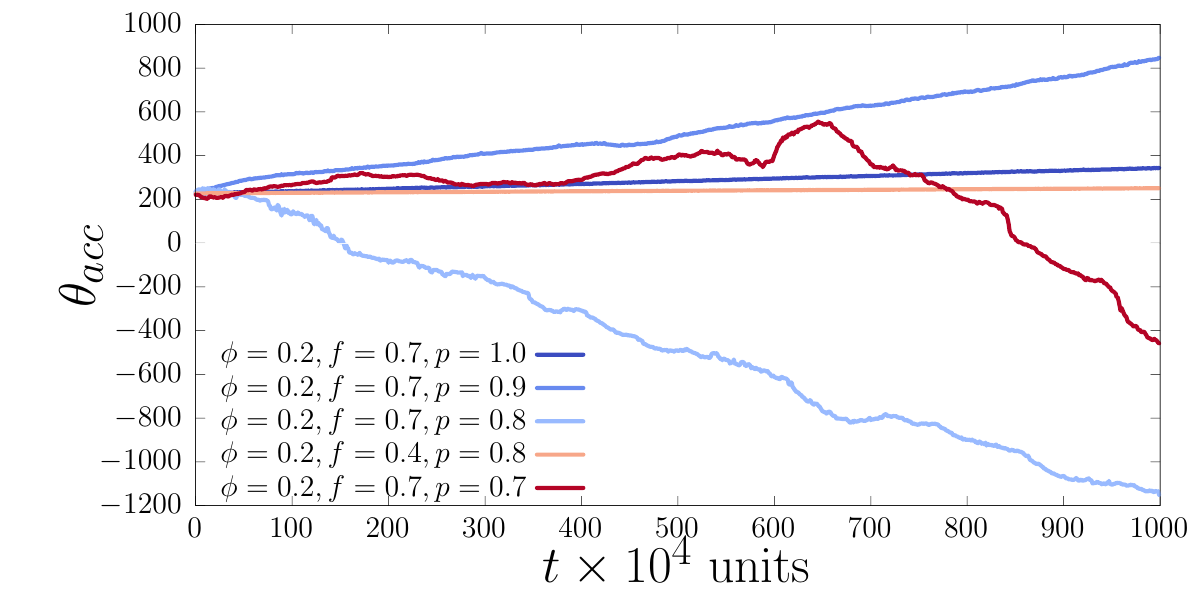}
\caption{\textbf{Rotation under the action of force dipoles:} Accumulated angle $(\theta_{acc}^{\degree})$ vs. time $(t)$ for the corner-most node about the center of mass of a $20\times 20$ nodes system with various degree of bond dilution $p=1, 0.9, 0.8$ and $0.7$. $\theta_{acc}^{\degree}>0$ denotes anti-clockwise, and $\theta_{acc}^{\degree}<0$ denotes clockwise rotations.}
\label{fig_rotation}
\end{figure}

To observe the bulk rotation of the lattice, we calculate numerically, the accumulated angle $(\theta_{acc})$ of a corner node about the center of mass, for a system of $20\times 20$ nodes, shown in Fig. \ref{fig_rotation}. The quantity $\frac{\theta_{acc}}{360\degree}$ then gives the number of complete rotations and $\Omega_i = \frac{\partial \theta_{acc}}{\partial t}$. We found the rotational speed to depend up on how far the system is placed from $p_{iso}$. For $p=1$ and $\phi=0.2$, the system shows unidirectional (clockwise or anti-clockwise) persistent rotation. Then, as we dilute the system $(p<1)$, the system starts showing tendency to shrink, as a result the angular velocity increases, e.g., for $p=0.9$ and $0.8$ [see Fig. \ref{fig_rotation}]. On further diluting the system down to $p\approx p_{iso}$, e.g., $p=0.7$, the system is almost collapsed, and display frequent changes in rotational direction, where the effective $\langle \theta_{ac} \rangle\approx 0$ over a large time. 

Angular velocity also depends on other factors like the force amplitude $(f)$, shown by comparing $f=0.4$ and $0.7$ for $p=0.8$, and also on the fraction of dipole links $(\phi)$, shown by comparing $\phi=0.2$ and $0.3$ for $p=1$, which results in the overall shrinkage of the network. However, for $\phi=1$ in $p=1$ lattices, we do not find any rotation, but only a fluctuation about a mean value in the steady state.

It is important to keep in mind that, such rotations essentially happens due to the residual anisotropy of the random realization of the dipolar forces, where the direction actually depends on the realization of force dipoles that slightly breaks the symmetry. Therefore, for larger systems the rotation should be slow or equivalently the angular velocity goes to zero. 

\section{Discussion and conclusion \label{conclusion}}

In conclusion, we studied the dynamic behavior of critical percolation clusters as a model for disordered fractal networks in the presence of active noise, which drives the system out of equilibrium, extending previous simulations on deterministic fractals. The active noise is modeled as a temporally correlated telegraphic process. We study bond percolation critical infinite clusters of square and triangular lattices. Three distinct cases are examined: (i) only thermal noise, (ii) active monopole forces, and (iii) active force dipoles. In a nutshell, the spectral dimension of a fractal is solely responsible for the sub-diffusion in the presence of either thermal fluctuations or active force monopoles. Sub-diffusion, however, is absent for active force dipoles, which is replaced by a saturation regime. Moreover, the critical percolation clusters always slowly collapse under force dipoles since they are below the so-called ``rigidity percolation" threshold. Importantly, the dynamics is only sensitive to one of the fractal dimensions viz. the spectral dimension $(d_s)$ and is independent of the other fractal dimensions such as $d_f$ and $d_l$. 

We extend the regime of study to undiluted and sparsely diluted systems as well in an endeavor to understand certain aspects (e.g., the collapse transition), which was present in deterministic fractal lattice (e.g., the Sierpinski gasket \cite{singh2024}), but is absent in critical percolation clusters as they collapse already at even tiny forces. A sharp collapse transition is observed only for a fully connected triangular lattice, which gradually smoothens as we dilute the system down to the rigidity percolation point. We found an extended rigid regime of the dipolar networks poised below the common rigidity percolation threshold of passive networks, due to the additional constraints of the active dipole links. This is similar to the behavior of pre-stressed networks \cite{alexander1998} whose rigidity is increased by the internal stresses.

The rotational motion is also found to be sensitive to the rigidity threshold. Fully connected triangular lattices show persistent unidirectional `` crawling" rotations --- i.e. a rotational swimming behavior -- similar to the Sierpinski gasket \cite{singh2024}. On weak dilutions, the mean angular velocity increases, possibly because of the increase of anisotropy of the force-dipole distribution. On stronger dilutions, however, as we approach the rigidity threshold, we observe increase in the angular velocity fluctuations, to the extent that we cannot assign anymore a mean value for it within the simulation time window. We conjecture that the mean angular velocity diminishes on approaching the rigidity percolation, while the increase of rotational fluctuations is a signature of critical dynamics, similar to equilibrium fluctuations near criticality; yet more studies are required to study these issues.

Our findings are crucial for understanding systems like chromatin dynamics, which exhibit chromosomal loci sub-diffusion in both normal and ATP-depleted cells with identical exponents \cite{weber2012}, similar to our theoretical findings.  They also show large-scale coherent motion, claimed to be associated with the action of force dipoles \cite{bruinsma20014,saintillan2018}. The latter could be associated with such rotational motion, and possibly longer range interactions and bending rigidity (absent in our study) substantially lower the rigidity threshold such that these coherent rotations are possible. Another example of rotational motion is the acto-myosin intracellular cortex that show persistent rotational motion \cite{woodhouse2012,suzuki2017}.

Finally, it is important to note that our research is of a much broader scope -- not limited to biological systems -- and concerns ``active networks" in general. While we are not simulating a specific physical system, by considering networks we have drawn some general conclusions that can help to understand a variety of systems with embedded active forces.

\section{Acknowledgement}
DM was supported by the BCSC Fellowship from the Jacob Blaustein Center for Scientific Cooperation and by the Israel Science Foundation (ISF) through Grant No. 1301/17 and 1204/23. SS acknowledges the BGU Kreitmann School postdoctoral fellowship. The authors thank the BGU Avram and Stella Goldstein-Goren Fund for support.


\begin{thebibliography}{0}%
\makeatletter
\providecommand \@ifxundefined [1]{%
 \@ifx{#1\undefined}
}%
\providecommand \@ifnum [1]{%
 \ifnum #1\expandafter \@firstoftwo
 \else \expandafter \@secondoftwo
 \fi
}%
\providecommand \@ifx [1]{%
 \ifx #1\expandafter \@firstoftwo
 \else \expandafter \@secondoftwo
 \fi
}%
\providecommand \natexlab [1]{#1}%
\providecommand \enquote  [1]{``#1''}%
\providecommand \bibnamefont  [1]{#1}%
\providecommand \bibfnamefont [1]{#1}%
\providecommand \citenamefont [1]{#1}%
\providecommand \href@noop [0]{\@secondoftwo}%
\providecommand \href [0]{\begingroup \@sanitize@url \@href}%
\providecommand \@href[1]{\@@startlink{#1}\@@href}%
\providecommand \@@href[1]{\endgroup#1\@@endlink}%
\providecommand \@sanitize@url [0]{\catcode `\\12\catcode `\$12\catcode
  `\&12\catcode `\#12\catcode `\^12\catcode `\_12\catcode `\%12\relax}%
\providecommand \@@startlink[1]{}%
\providecommand \@@endlink[0]{}%
\providecommand \url  [0]{\begingroup\@sanitize@url \@url }%
\providecommand \@url [1]{\endgroup\@href {#1}{\urlprefix }}%
\providecommand \urlprefix  [0]{URL }%
\providecommand \Eprint [0]{\href }%
\providecommand \doibase [0]{http://dx.doi.org/}%
\providecommand \selectlanguage [0]{\@gobble}%
\providecommand \bibinfo  [0]{\@secondoftwo}%
\providecommand \bibfield  [0]{\@secondoftwo}%
\providecommand \translation [1]{[#1]}%
\providecommand \BibitemOpen [0]{}%
\providecommand \bibitemStop [0]{}%
\providecommand \bibitemNoStop [0]{.\EOS\space}%
\providecommand \EOS [0]{\spacefactor3000\relax}%
\providecommand \BibitemShut  [1]{\csname bibitem#1\endcsname}%
\let\auto@bib@innerbib\@empty
\end{thebibliography}%


\begin{thebibliography}{40}
\providecommand{\natexlab}[1]{#1}
\providecommand{\url}[1]{\texttt{#1}}
\expandafter\ifx\csname urlstyle\endcsname\relax
  \providecommand{\doi}[1]{doi: #1}\else
  \providecommand{\doi}{doi: \begingroup \urlstyle{rm}\Url}\fi

\bibitem[MacKintosh and Levine(2008)]{mac2008}
F~C MacKintosh and A~J Levine.
\newblock Nonequilibrium mechanics and dynamics of motor-activated gels.
\newblock \emph{Physical Review Letters}, 100\penalty0 (018104), 2008.

\bibitem[Alvarado et~al.(2013)Alvarado, Sheinman, Sharma, MacKintosh, and Koenderink]{alvarado2013}
Jose Alvarado, Michael Sheinman, Abhinav Sharma, Fred~C MacKintosh, and Gijsje~H Koenderink.
\newblock Molecular motors robustly drive active gels to a critically connected state.
\newblock \emph{Nature Physics}, 9\penalty0 (11), 2013.

\bibitem[Prost et~al.(2015)Prost, Julicher, and Joanny]{prost2015}
J~Prost, F~Julicher, and J-F Joanny.
\newblock Active gel physics.
\newblock \emph{Nature Physics}, 11:\penalty0 111--117, 2015.

\bibitem[Alvarado et~al.(2017)Alvarado, Sheinman, Sharma, MacKintosh, and Koenderink]{alvarado2017}
Jose Alvarado, Michael Sheinman, Abhinav Sharma, Fred~C MacKintosh, and Gijsje~H Koenderink.
\newblock Force percolation of contractile active gels.
\newblock \emph{Soft Matter}, 13\penalty0 (5624), 2017.

\bibitem[Broedersz et~al.(2011)Broedersz, Mao, Lubensky, and MacKintosh]{broedersz2011}
C~P Broedersz, Xiaoming Mao, T~C Lubensky, and F~C MacKintosh.
\newblock Criticality and isostaticity in fibre networks.
\newblock \emph{Nature Physics}, 7\penalty0 (983-988), 2011.

\bibitem[Kumar et~al.(2023)Kumar, Quint, and Dasbiswas]{dasbiswas2023}
Abhinav Kumar, David~A Quint, and Kinjal Dasbiswas.
\newblock Range and strength of mechanical interactions of force dipoles in elastic fiber networks.
\newblock \emph{Soft Matter}, 19:\penalty0 5805--5823, 2023.

\bibitem[Metze(2013)]{metze2013}
K~Metze.
\newblock Fractal dimension of chromatin: potential molecular diagnostic applications for cancer prognosis.
\newblock \emph{Expert Review of Molecular Diagnostics}, 13\penalty0 (7):\penalty0 719--735, 2013.

\bibitem[Gnesotto et~al.(2019)Gnesotto, Remlein, and Broedersz]{gnesotto2019}
Federico~S Gnesotto, B~M Remlein, and C~P Broedersz.
\newblock Nonequilibrium dynamics of isostatic spring networks.
\newblock \emph{Phys. Rev. E}, 100\penalty0 (013002), 2019.

\bibitem[Jurjiu et~al.(2003)Jurjiu, Koslowski, and Blumen]{jurjiu2003}
A~Jurjiu, Th. Koslowski, and A~Blumen.
\newblock Dynamics of deterministic fractal polymer networks: Hydrodynamic interactions and the absence of scaling.
\newblock \emph{J. Chem. Phys.}, 118:\penalty0 2398--2404, 2003.

\bibitem[Jurjiu and Galiceanu(2018)]{jurjiu2018}
Aurel Jurjiu and Mircea Galiceanu.
\newblock Dynamics of a polymer network modeled by a fractal cactus.
\newblock \emph{Polymers}, 10\penalty0 (787), 2018.

\bibitem[Grosberg et~al.(1993)Grosberg, Rabin, Havlin, and Neer]{grosberg1993}
A~Grosberg, Y~Rabin, S~Havlin, and A~Neer.
\newblock Crumpled globule model of the three-dimensional structure of dna.
\newblock \emph{Europhysics Letters}, 23\penalty0 (5):\penalty0 373--378, 1993.

\bibitem[Tamm et~al.(2015)Tamm, Nazarov, Gavrilov, and Chertovich]{tamm2015}
M~V Tamm, L~I Nazarov, A~A Gavrilov, and A~V Chertovich.
\newblock Anomalous diffusion in fractal globules.
\newblock \emph{Phys. Rev. Lett.}, 114\penalty0 (178102), 2015.

\bibitem[Singh and Granek(2024)]{singh2024}
Sadhana Singh and Rony Granek.
\newblock Active fractal networks with stochastic force monopoles and force dipoles unravel subdiffusion of chromosomal loci.
\newblock \emph{Chaos}, 34\penalty0 (113107), 2024.

\bibitem[Woodhouse and Goldstein(2012)]{woodhouse2012}
F.~G Woodhouse and R.~E. Goldstein.
\newblock Spontaneous circulation of confined active suspensions.
\newblock \emph{Phys. Rev. Lett.}, 109\penalty0 (168105), 2012.

\bibitem[Suzuki et~al.(2017)Suzuki, Miyazaki, Takagi, Itabashi, and Ishiwata]{suzuki2017}
Kazuya Suzuki, Makito Miyazaki, Jun Takagi, Takeshi Itabashi, and Shin'ichi Ishiwata.
\newblock Spatial confinement of active microtubule networks induces large-scale rotational cytoplasmic flow.
\newblock \emph{Proceedings of the National Academy of Sciences}, 114\penalty0 (11):\penalty0 2922--2927, 2017.

\bibitem[Kumar et~al.(2014)Kumar, Sumit, Ramaswamy, and Shivashankar]{kumar2014}
Maitra~Ananyo Kumar, Abhishek, Madhuresh Sumit, Sriram Ramaswamy, and G~V Shivashankar.
\newblock Actomyosin contractility rotates the cell nucleus.
\newblock \emph{Scientific Reports}, 4\penalty0 (3781), 2014.

\bibitem[Stauffer(1992)]{stauffer1992}
Dietrich Stauffer.
\newblock \emph{Introduction to percolation theory}.
\newblock Taylor \& Francis, London, 2nd edition, 1992.

\bibitem[Alexander and Orbach(1982)]{alex1982}
S~Alexander and R~Orbach.
\newblock Density of states on fractals : fractons.
\newblock \emph{J. Physique Lettres}, 43:\penalty0 L--625 -- L--631, 1982.

\bibitem[Alexander(1989)]{alex1989}
S~Alexander.
\newblock Vibrations of fractals and scattering of light from aerogels.
\newblock \emph{Phys. Rev. B}, 40\penalty0 (11), 1989.

\bibitem[Gardiner(1991)]{gardiner1991}
C~Gardiner.
\newblock \emph{Handbook of Stochastic Processes, and Its Applications to Physics, Chemistry and the Natural Sciences, Springer Series in Synergetics}.
\newblock Springer, New York, 1991.

\bibitem[Hoshen and Kopelman(1976)]{hoshen1976}
J~Hoshen and R~Kopelman.
\newblock Percolation and cluster distribution. i. cluster multiple labeling technique and critical concentration algorithm~.
\newblock \emph{Physical Review B}, 14\penalty0 (8):\penalty0 3438--3445, 1976.

\bibitem[Granek and Klafter(2005)]{granek2005}
Rony Granek and Joseph Klafter.
\newblock Fractons in proteins: Can they lead to anomalously decaying time autocorrelations?
\newblock \emph{Physical Review Letters}, 95\penalty0 (098106), 2005.

\bibitem[Granek(2011)]{granek2011}
Rony Granek.
\newblock Proteins as fractals: Role of the hydrodynamic interaction.
\newblock \emph{Physical Review E}, 83\penalty0 (020902(R)), 2011.

\bibitem[Reuveni et~al.(2012{\natexlab{a}})Reuveni, Klafter, and Granek]{reuveni2012}
Shlomi Reuveni, Joseph Klafter, and Rony Granek.
\newblock Dynamic structure factor of vibrating fractals: Proteins as a case study.
\newblock \emph{Physical Review E}, 85\penalty0 (011906), 2012{\natexlab{a}}.

\bibitem[Reuveni et~al.(2012{\natexlab{b}})Reuveni, Klafter, and Granek]{reuveni2012p2}
Shlomi Reuveni, Joseph Klafter, and Rony Granek.
\newblock Dynamic structure factor of vibrating fractals.
\newblock \emph{Phys. Rev. Lett.}, 108\penalty0 (068101), 2012{\natexlab{b}}.

\bibitem[to~the Maxwell's rule for mechanical stability in a $d$-dimensional network at the isostatic point the mean coordination number $\langle z \rangle$ should be at least $\langle z \rangle=2d$ which is 4 for two-dimensional systems()]{comm1}
According to~the Maxwell's rule for mechanical stability in a $d$-dimensional network at the isostatic point the mean coordination number $\langle z \rangle$ should be at least $\langle z \rangle=2d$ which is 4 for two-dimensional systems.

\bibitem[Das et~al.(2012)Das, Quint, and Schwarz]{das2012}
Moumita Das, David~A Quint, and J~M Schwarz.
\newblock Redundancy and cooperativity in the mechanics of compositely crosslinked filamentous networks.
\newblock \emph{PLoS ONE}, 7(5):e35939, 2012.

\bibitem[ALEXANDER(1998)]{alexander1998}
Shlomo ALEXANDER.
\newblock Amorphous solids: Their structure, lattice dynamics and elasticity.
\newblock \emph{Physics Reports}, 296:\penalty0 65--236, 1998.

\bibitem[Weber et~al.(2011)Weber, Spakowitz, and Theriot]{weber2012}
Stephanie~C Weber, Andrew~J Spakowitz, and Julie~A Theriot.
\newblock Nonthermal atp-dependent fluctuations contribute to the in vivo motion of chromosomal loci.
\newblock \emph{Proceedings of the National Academy of Sciences}, 109\penalty0 (19):\penalty0 7338--7343, 2011.

\bibitem[Bruinsma et~al.(2014)Bruinsma, Grosberg, Rabin, and Zidovska]{bruinsma20014}
R~Bruinsma, A~Y Grosberg, Y~Rabin, and A~Zidovska.
\newblock Chromatin hydrodynamics.
\newblock \emph{Biophysical Journal}, 106:\penalty0 1871, 2014.

\bibitem[Saintillan et~al.(2018)Saintillan, Shelley, and Zidovska]{saintillan2018}
D~Saintillan, M~J Shelley, and A~Zidovska.
\newblock Extensile motor activity drives coherent motions in a model of interphase chromatin.
\newblock \emph{Proc. Natl. Acad. Sci.}, 115\penalty0 (11442-11447), 2018.

\bibitem[Cui and Bustamante(2000)]{cui2000}
Y~Cui and C~Bustamante.
\newblock \emph{Proceedings of the National Academy of Sciences}, 97\penalty0 (127), 2000.

\bibitem[Johnstone et~al.(2020)Johnstone, Wang, Sevier, and Galloway]{johnstone2020}
C~P Johnstone, N~B Wang, S~A Sevier, and K~E Galloway.
\newblock \emph{Cell Systems}, 11\penalty0 (424), 2020.

\bibitem[Fierz and Poirier(2019)]{fierz2019}
B~Fierz and M~G Poirier.
\newblock \emph{Annual review of Biophysics}, 48\penalty0 (321), 2019.

\bibitem[Zhou et~al.(2016)Zhou, Johnson, Gamarra, and Narlikar]{zhou2016}
C~Y Zhou, S~L Johnson, N~I Gamarra, and G~J Narlikar.
\newblock \emph{Annual review of Biophysics}, 45\penalty0 (153), 2016.

\bibitem[MacKintosh et~al.(1995)MacKintosh, Kas, and Janmey]{mackintosh1995}
F~MacKintosh, J~Kas, and P~Janmey.
\newblock \emph{Phys. Rev. Lett.}, 75:\penalty0 4425, 1995.

\bibitem[Broedersz and MacKintosh(2014)]{broedersz2014}
C~P Broedersz and F~C MacKintosh.
\newblock \emph{Rev. Mod. Phys.}, 86:\penalty0 995, 2014.

\bibitem[Wedemann and Langowski(2002)]{wedemann2002}
G~Wedemann and J~Langowski.
\newblock \emph{Biophysical Journal}, 82:\penalty0 2847, 2002.

\bibitem[Ghosh and Gov(2014)]{ghosh2014}
A~Ghosh and N~Gov.
\newblock \emph{Biophysical Journal}, 107:\penalty0 1065, 2014.

\bibitem[Deviri et~al.(2017)Deviri, Discher, and Safran]{deviri2017}
D~Deviri, D.~E. Discher, and S.~A. Safran.
\newblock \emph{Biophysical Journal}, 113:\penalty0 1060, 2017.

\end{thebibliography}


\appendix

\section{Simulation and Methods}

\subsection{Numerical integration  \label{simulation_appendix}}

To numerically integrate the ordinary differential equations (ODEs) of motion we use the two-point finite difference method, also known as the second order Heun algorithm or the midpoint scheme for ODEs. This consists of the conventional Euler-Maruyama scheme (Eq. \ref{euler}(a)) along with an additional corrector step (Eq. \ref{euler}(b)) as mentioned below 
\begin{subequations}
\begin{eqnarray}
\bar{\vec r}_{k+1} = \vec r_k + \frac{1}{\gamma}F (\vec r_k) \delta t  + \sqrt{\frac{2k_BT}{\gamma}} \vec W_k   \\
\vec r_{k+1} = \vec r_k + \frac{1}{2\gamma} \left( F (\vec r_k) + F (\bar{\vec r}_{k+1}) \right) \delta t \nonumber \\
+ \sqrt{\frac{2k_BT}{\gamma}} \vec W_k 
\end{eqnarray}
\label{euler}
\end{subequations}
where $\vec F (\vec r_k)$ are the total elastic forces and $\sqrt{\frac{2k_BT}{\gamma}} \vec W_k$ are normally distributed random vectors. In this integration scheme, a predicted position $\bar{\vec r}_{k+1}$ (Eq. \ref{euler}a) is followed by a corrected position $\vec r_{k+1}$ (Eq. \ref{euler}b). In both the steps, the same realization of random vectors $\vec W_k$ are used as thermal noise. For active systems, the above equations are modified using additional terms for active forces. 

\subsection{Telegraphic noise \label{tlnoise}}

 The source of active forces (or noise) switches between the $0$ or $f$ state following the transition rules as mentioned below,
\begin{eqnarray}
 \mathcal{R} &<&  (1-\mathcal{P})+\mathcal{P}\times \exp(-\delta t/\tau),~~~~ 0\rightarrow 0 \\
 \mathcal{R} &<& \mathcal{P}+(1-\mathcal{P})\times \exp(-\delta t/\tau),~~~~ f\rightarrow f,
\end{eqnarray} 
where $\mathcal{R}$ is a uniformly distributed random number $\mathcal{R}\in [0,1]$, $\mathcal{P}$ is the probability of a dipole source to be ON initially at $t=0$, $\tau$ is the decorrelation time, and $\delta t$ is the integration time step.

\subsection{Parameter choice \label{parameter_choice}}

To relate our findings to the biological systems, we use the parameter values in connection to the cytoplasmic active processes as described in Ref. \cite{singh2024}. We assume $\tau$ and $f$ to be associated with the cytoplasmic motor protein processivity times and forces. 
In cytoplasm, typical values are in the range $f_0=1-10$pN and $\tau=10^{-3}-10$s  \cite{cui2000, johnstone2020, fierz2019, zhou2016}. Since, the network segment between two beads behaves as a Gaussian 2D thermal spring with spring constant 
\begin{equation}
m \omega_0^2 = \frac{90k_BT L_p^2}{b^4}
\end{equation}
for semi-flexible polymers \cite{mackintosh1995, broedersz2014}, where the bond length between two beads $(b)$ acts as the persistence length $b=L_p\approx 250$nm \cite{wedemann2002, ghosh2014, deviri2017}. Assuming the solvent (water) viscosity $\eta=10^{-3}$Pa, the normalization factors can be estimated to be 
\begin{equation}
\text{Unit of force} = m \omega_0^2 b = \frac{90 k_B T}{b} = 1.5\text{pN}
\end{equation}
using $k_BT=4.1$pN nm, and
\begin{equation}
\tau_0 = \frac{\gamma}{m \omega_0^2}=\frac{3\pi\eta b^2}{90k_BT}=4\times 10^{-4}.
\end{equation}
Therefore, the dimensionless value of the force amplitude is in the range $f_0=0.5-7$, and the force decorrelation time in the range $\tau=1-10^4$. 

\section{Modified Maxwell criterion \label{mmc}}

For a two-dimensional network, where each node has 2 degrees of freedom (d.o.f),
\begin{equation}
    \text{total number of d.o.f} =2N.
\end{equation}
Excluding the d.o.f. due to the rigid body motions; 2 translational and 1 rotational d.o.f, we have 
\begin{equation}
    \text{total number of d.o.f} =2N-3.
\end{equation}
Now, each link between a pair of nodes presents a constraint. Let's say there are $M$ links (or edges) in the system. Thereafter, for the system to be just rigid, the total number of independent d.o.f must equal the number of independent constraints, i.e.,  
\begin{equation}
    M=2N-3.
    \label{eqm}
\end{equation}
Let's say $z$ is the coordination number, then assuming periodic boundary condition, $M=zN/2$, where the $1/2$ factor removes double counting of links. Therefore, from Eq. \ref{eqm} we have, 
\begin{equation}
    \frac{z_cN}{2} = 2N-3
    \label{eqzc}
\end{equation}
which in the thermodynamic limit becomes 
\begin{equation}
    \lim_{N\rightarrow \infty} z_c=4.
\end{equation}

Therefore, the probability $p$ with which a link should exist for the system to be just rigid would be 
\begin{equation}
    p=\frac{z_c}{z}=\frac{4}{6}=\frac{2}{3}
\end{equation}
which is the isostatic point in 2D.

When stochastic force dipoles act on a fraction $\phi$ of links, with a fraction $\mathcal{P}$ out of them in their ON state, this implies an additional force $f$ on $\mathcal{P}\phi$ fraction of links. Hence, the number of independent constraints will change to $M + \mathcal{P}\phi M$. Therefore, from Eq. \ref{eqzc} we have 
\begin{equation}
    \frac{z'_cN}{2}(1+\mathcal{P}\phi) = 2N-3,
\end{equation}
which in the thermodynamic limit becomes
\begin{equation}
    \lim_{N\rightarrow \infty} z'_c=\frac{4}{1+\mathcal{P}\phi}.
\end{equation}
Therefore, the probability with which a link should exist for the system to be just rigid 

\begin{equation}
    p' = \frac{z'_c}{z}=\frac{z_c}{z(1+\mathcal{P}\phi)} = \frac{2}{3(1+\mathcal{P}\phi)}.
\end{equation}
For example, for $\phi=0.2$, new critical point for the system to become rigid would be $p'=0.606$, which matches well with our simulations [see Fig. \ref{fig_collapse}(c)]. However, for finite size lattices, we expect a deviation $\mathcal{O}(1/\sqrt{N})$.

\end{document}